\documentclass[pre, twocolumn, showpacs, showkeys]{revtex4}
\usepackage{epsfig}
\begin{document}
\draft
\title{Instant Messaging as a Scale-Free Network}
\author{Reginald Smith}
\affiliation{University of Virginia}
\email{rds2u@alumni.virginia.edu}
\today

\begin{abstract}
The topology of an instant messaging system is described. Statistical measures
of the network are given and compared with the statistics of a comparable
random graph. The scale-free character of the network is examined and
implications are given for the structure of social networks and instant
messenger security.
\end{abstract}

\address{P.O. Box 360262 Decatur, GA 30036}
\maketitle
\section{Introduction}

The last few years has seen a large advance in our understanding of networks
whose structures are by nature non-equilibrium and non-random. These networks
have been used to study systems as diverse as the Internet router and WWW
hyperlink networks, electric power grids, and cellular metabolic pathways
\cite{review1, review2}. In particular, these networks prominently feature a
power-law frequency distribution for the nodes' degree (scale-free network), a
network diameter smaller than a comparable random graph, one with the same
amount of nodes and the same average degree per node, and a much larger
clustering coefficient than a comparable random graph.

Among the most interesting of the networks studied are those that analyze
human social interaction. The phenomenon of six-degrees of separation, first
recognized by Stanley Milgram\cite{milgram}, is well known and documented in
both academic and popular culture. The most detailed of these networks studied
are actor collaboration networks and scientific collaboration
networks\cite{actor1, actor2, sciencecollab1, sciencecollab2}. Interestingly,
the web of human sexual partners have also been documented\cite{lottasex}.
Typically these networks share the features mentioned above that differentiate
them from random graphs and display a scale-free character. This paper hopes to
add to these studies on social contact networks by adding another example: the
connections of users in an instant messaging service.

\section{Overview}

Instant messaging has grown at a phenomenal rate in the last several years to
become a major form of communication both over the internet and within company
intranets. Instant messaging has become so important in fact that the FCC
has attempted to force the largest server for instant messaging, AOL
Time-Warner, to open its software for interoperability
\cite{FCC}.

Instant messaging is distinguished from regular chat as being a one-on-one
conversation between two users on an instant messaging network. Typically in
instant messaging systems each user has a user name and a contact list
containing the user names of other users who they often communicate with. It
is this feature of instant messaging that makes it amenable to scientific
study and statistical analysis. If one imagines each user as a node and each
contact on ther user's contact list as an out-directed edge, the community on
an instant messaging network can be modeled using graph theory.

Using these assumptions it can be easily seen that an instant messaging
network represents a non-equilibrium graph in that nodes (users) are added and
removed over time and edges most likely accumulate on users in a non-random
fashion. One possible model for this growth is the Barab\'{a}si-Albert (BA)
model where edges are formed by preferential attachment. Those nodes with more
edges are more likely to accumulate edges as time goes on. Similarly one could
hypothesize a user is more likely to form out-directed edges (add users to the
user's contact list) the more users are already present on their contact list
and a user is more likely to receive in-directed edges (being on another
user's contact list) in a similar fashion.

With these assumptions an instant messaging network's graph should probably
display a scale-free character. To test this hypothesis an instant messaging
network using the open-source Jabber protocol was researched to find such
features.

\subsection{The Jabber Instant Messaging Protocol}

To clearly understand some of the assumptions and conclusions in this paper a
cursory overview of the Jabber protocol is necessary \cite{jabber}. The Jabber
protocol is based off of XML and uses a distributed client-server architecture.
Jabber was consciously based off of the architecture of email systems so
instead of one central server like AOL's Instant Messenger or Microsoft's MSN
Messenger, Jabber has many servers in many locations. Jabber clients are the
users who communicate with instant messaging. Clients can communicate with all
other clients on their Jabber servers and with clients on other Jabber servers
since the Jabber servers can communicate with each other. In addition Jabber
supports additions called transports that allow Jabber clients to communicate
with clients using other protocols such as those on AOL, Microsoft, ICQ, or
Yahoo instant messaging.

\section{Network Statistics}

The network studied was the instant messaging database from nioki.com, a
French language teen-oriented web site. Appropriate measures were taken to
completely preserve the anonymity and privacy of the users as is explained in
detail in Appendix A. The nioki.com database contained 50,158 users (nodes)
with almost 500,000 edges. Due to the model explained earlier, this instant
messaging network was modeled as a directed graph. This is different from
other social network studies such as the actor-movie collaboration network
which was modeled as a bipartite graph and the scientific collaborations
which were modeled as undirected graphs.

The nioki.com instant messaging network was found to exhibit all the
characteristics of a scale free network. The inward and outward directed edge
frequency distributions both followed power laws with a $\gamma_{in} =$ 2.2
 and $ \gamma_{out} =$ 2.4. The average in and out degrees are $\langle k_{in}
\rangle =$ 9.1  and $ \langle k_{out} \rangle =$ 8.2. The average in and out
degrees are identical in a network with no outside contacts. However, as
explained in the description of Jabber, clients have the ability to
communicate with clients on other servers outside their current server. So
there are probably contacts with clients that are not on nioki.com's server.
The data does not indicate who these contacts are but the difference in the
average in and out degrees per node intimate their existence. The diameter of
the network, $\overline{\ell}$ = 4.35 so that there are about 4-5 users on
average between any two users on the network. These values indicate the small
world character of the nioki.com network as compared to a random graph (Table
\ref{tab:compare}). 

Since this network is modeled as a directed graph, it presents a rather
asymmetric view of human social interaction. In a directed graph it is
possible for one user to "know" another without the other user reciprocally
expressing such a relationship. This is because the contact list data we have
does not require a reciprocal relationship between two users. Measurements
indicate about 82\% of the contacts in the network are in both directions. So
on average 82\% of the users on a given user's (user A) contact list also have
the user A on their contact list. In order to get a clearer view and calculate
the clustering coefficient, a new list of users and contacts was created
adding those edges necessary to make the network undirected. In this case the
power exponent of the degree distribution became $\gamma = 1.8$. The average
degree was $\langle k \rangle =$ 9.6 and  the diameter of the network decreased to
$\overline{\ell}$ = 4.1. The average clustering coefficient was calculated at
$\overline{C}$ = 0.33 further reinforcing the small-world character of the
network. 

The final measures computed were the size of the giant weakly connecting
connected component (GWCC) and the giant strongly connected component (GSCC).
The GWCC is the number of nodes on the network
that can be reached by any other node in the component ignoring the directions
of edges. It was calculated at a very large 49,801 users or over 99\% of the
users on nioki.com. Only about 0.7\% of the users were in disconnected
components. The GSCC is a measure of the number of nodes in the component where
any node can be reached from any other node through directed edges. This was
calculated to be 44,581 or 89\% of nioki.com's users. These very large values
for the connected components are probably most likely explained by the
structure of the nioki.com website itself. It is mostly made up of users who
communicate using the nioki instant messaging service with other users on
nioki.com. It is unlikely that nioki.com's instant messenger is used as a
primary instant messaging tool for users on other servers or services by most
users. Being a teen oriented website geared toward socialization it likely has
a tightly knit community over shared interests. This is unlike the actor or
scientific collaboration databases where communication is mainly limited to
professional roles and fields of research. 

\begin{table}
\begin{tabular}[b]{|c||p{50pt}|p{50pt}|p{50pt}|}
\cline{1-4} &nioki.com  directed&nioki.com  undirected&random graph\\
\hline
No. of nodes&50158&50158&50158\\
$\langle k_{in} \rangle$ &9.1&9.6&9.6\\  
$\langle k_{out} \rangle$ &8.2&9.6&9.6\\
$\overline{C}$ &--&0.33&$1.9 x 10^{-4}$\\
$\overline{\ell}$& 4.35&4.1&4.79\\
\hline
\end{tabular}
\caption{\label{tab:compare}Comparison of Network Statistical Measures}
\end{table}

\begin{figure}

	\centering
	\includegraphics[height=2in, width=2in]{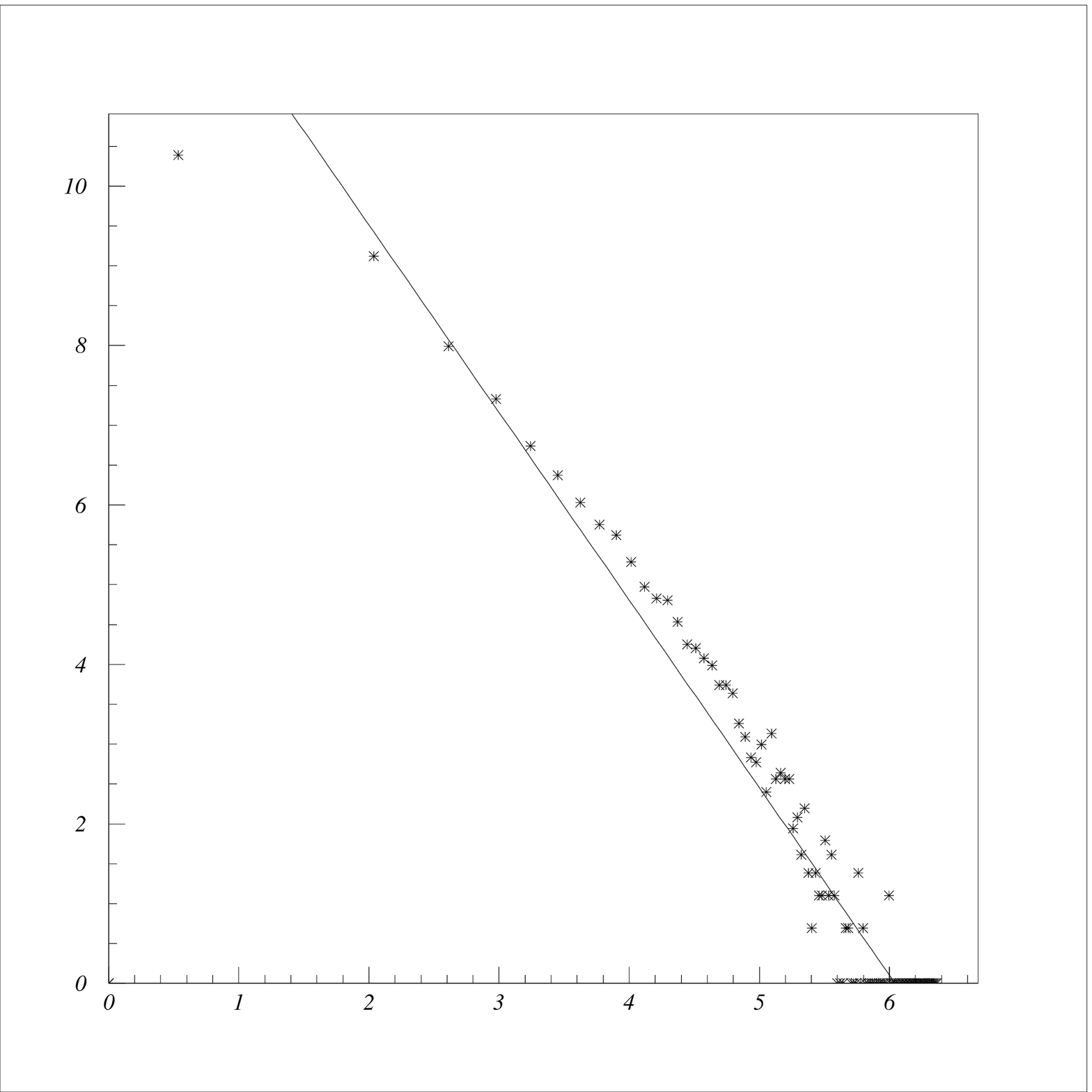}
        \caption{Log-log degree distribution of out-directed edges in the
         nioki.com instant messenger network}
  	
        \includegraphics[height=2in, width=2in]{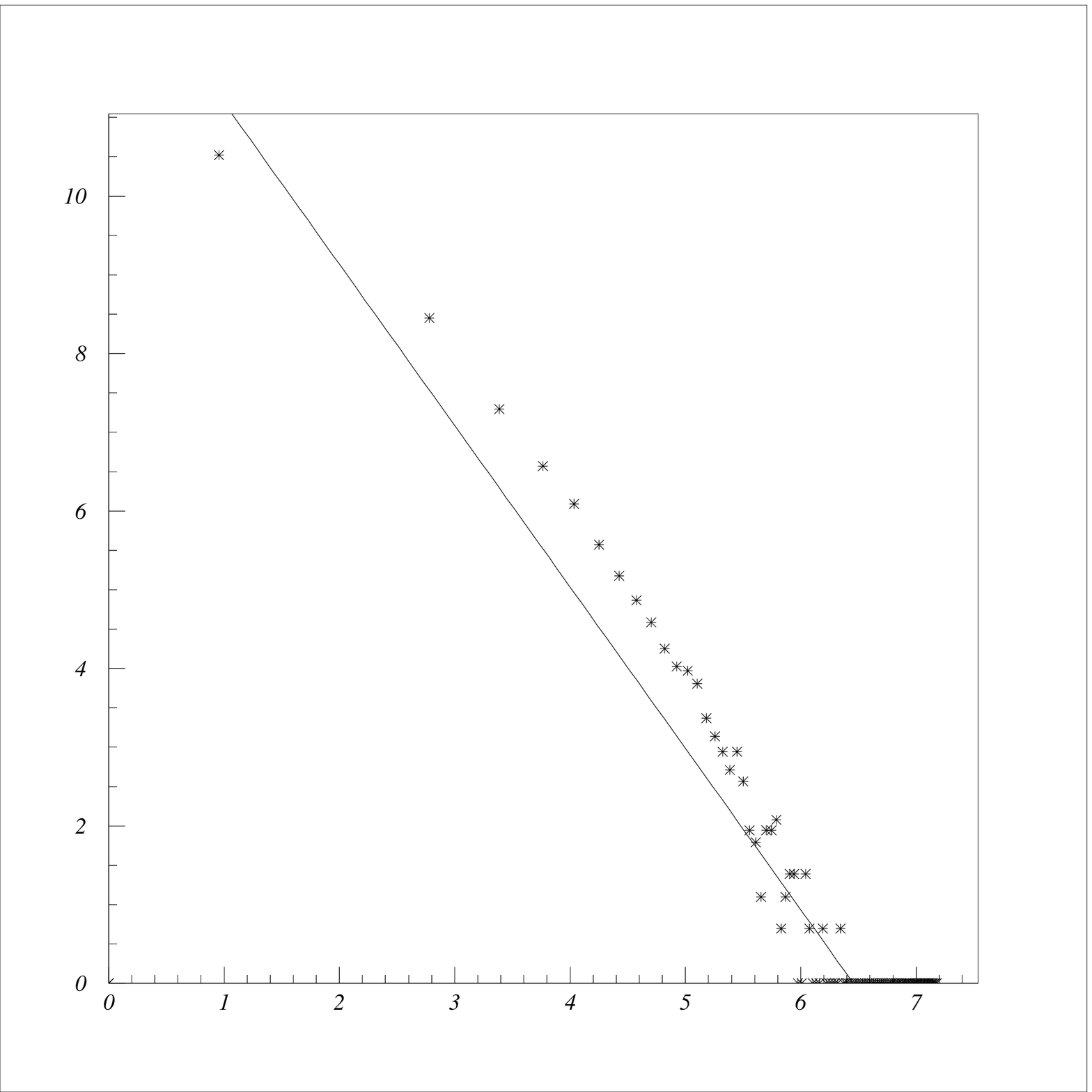}
        \caption{Log-log degree distribution of in-directed edges in the
          nioki.com instant messenger network}        
	\includegraphics[height=2in, width=2in]{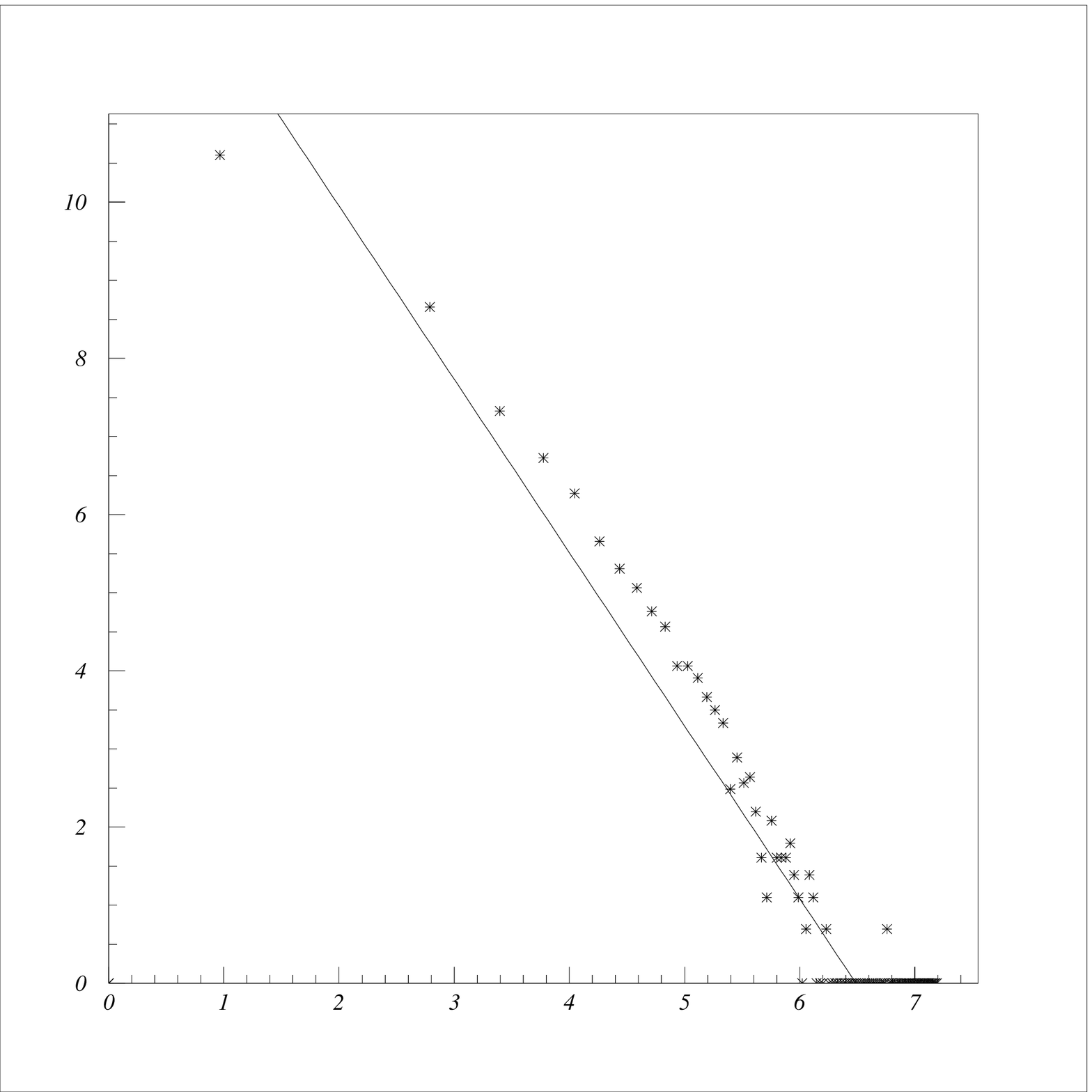}   
	\caption{Log-log degree distribution of constructed undirected
         nioki.com instant messenger network}

\end{figure}


\section{Calculations of random graph attributes}
The random graphs statistics in Table \ref{tab:compare} were computed using
the following equations which are theoretically explained in detail in
\cite{review1}. The comparable random graph was assumed to have the same
number of nodes and average degree per node as the undirected model of the
instant messenger network. From this information the average clustering
coefficient of the random graph can be calculated as \begin{equation}
\overline{C} = \frac{\langle k \rangle}{N} \end{equation} It is interesting to note that
the clustering coefficient of a random graph is the same as the node
connection probability. The shortest path was estimated using the
approximation \begin{equation}\overline{\ell} =
\frac{ln(N)}{ln(\langle k \rangle)}\end{equation}

 \section{Comparisons with Barab\'{a}si-Albert model} Earlier,
the possiblity that this network grows according to the Barab\'{a}si-Albert
model was considered. A key indication of this would be a measure of the
preferential attachment probability $\Pi(k)$. Though the empirical data from
the network and its scale-free character hint strongly towards the
Barabas\'{i}-Albert model or a comparable one, time dependent data was not
available to allow the determination of the shape (linear of curved) or
function of $\Pi(k)$.

\section{Relevance to Social Networks}
A frequent question with the ever faster globalization and communication in
the world is how connected we all really are. This research covered a
relatively large sample of about 50,000 people and in some ways gives a
glimpse into the connectivity of our society, but in other ways falls short.
\\
This research should give additional credence to the growing evidence of the
scale-free nature of social and professional contacts in greater society.
Combined with the earlier studies on professional collaborations it seems to
indicate that society does exhibit a "small world" effect. This research
finds that the small world of nioki.com is based on a scale-free topology,
however, there are other models of small worlds including the Watts-Strogatz
model which exhibit similar features. The small diameter compared to the
number of nodes in the network indicates that the degrees of separation in
nioki.com are a bit smaller than the six Stanley Milgram measured in his
studies. However, there are caveats to the wholesale application of these
results to larger society. Nioki.com is probably more connected than society
at large due to its foundation of shared interests and demographics (young
adults). The large size of the GWCC and GSCC is an indication of this. Thus,
the researcher does not think it would be completely accurate to say everyone
in the world is connected by 4-5 people. In fact in a recent criticism of
Milgram's work\cite{antimilgram}, it is asserted that studies that do not take
into account the increased likelihood of connections based on factors such
as demographics or professional affiliation may not clearly represent the
larger society. Though I believe this research further emphasizes that human
social networks have a small world character, this research cannot address the
question of the connectivity across varying demographics or social boundaries.

\section{Applications to Instant Messaging Security}

In instant messaging, like all computer network communications tools, security
is often a paramount consideration. Though there are many problems of interest
in the security of instant messaging from the privacy of conversations to the
security of user accounts and passwords the aspect of security most pertinent
here is the spread of worms across instant messenger networks. There have been
several recent outbreaks of worms on instant messaging networks \cite{worm1}.
The spread of epidemics on scale-free and small world networks has been well
studied. It is known that there is no epidemic threshold for
infinitely large scale-free networks \cite{spread1, spread2} and worms or viruses
can spread rapidly through a network. Though there have not been any
devastating worms so far it is wise to prepare to interdict the worst
possibility. Let us assume a worm spreads through an instant messaging network
by infecting a node and then spreading itself along all or some of the
out-directed edges to new nodes which also may be infected. This is similar to
recent worms which send a message containing an infected link to all members
of a user's contact list. The dynamics of this kind of epidemic have been
discussed \cite{spread1, spread2, spread3}. However, the options for stopping or slowing the
epidemic are varied. You can alert users or provide a patch or software to
prevent the spread as was done with the Code Red virus that infected Windows
2000 servers \cite{CAIDA}. With a more extreme event, however,
more radical measures could be necessary.

Unlike the Internet, where control is decentralized, the client-server nature
of instant messaging makes more radical measures possible. Research has
indicated that scale-free networks though they are robust to random node
failures, are very vulnerable to attack \cite{attack1, attack2, attack3}. A
directed attack at the most connected nodes could severly damage a network
such as the Internet. However, this characteristic could be reversed and
turned to an advantage in halting an epidemic \cite{AIDS, spread2}. In a severe instant
messaging worm outbreak the server administrators could slow, but not
completely stop, the spread of a worm in a way that does not affect the
service for many users. By disabling the accounts of the most connected users
on the network, they could effectively increase the network's diameter making
the propagation of the epidemic much slower and buying time for a patch or
another curative measure. The diameter of the nioki.com network after a
certain percentage of the most connected sites are removed is shown in Figure
4. Removing the top 10\% connected users increases the diameter of the network
almost twofold. However, even disabling up to 10\% of the most connected users
would leave connectivity for the other 90\% of the network and allow the
service to have more time to cope with the outbreak and help other users.
There are caveats to this plan, however. This would only work assuming that
the most connected users are online frequently enough to spread the worm. If
many of the most connected users are rarely online, this strategy may not
produce its full effect. Also, this strategy would still deny a segment of
users usage to the network for an unspecified period of time and would
possibly upset many users if it is used too frequently. 

\begin{figure}

	\centering
	\includegraphics[height=2in, width=2in]{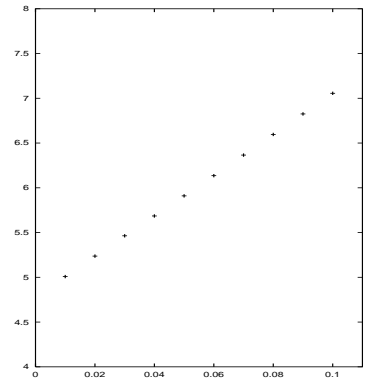}
        \caption{Graph of changes in the diameter of the nioki.com directed
network. Network diameter vs. percentage of most connected nodes removed} 
\end{figure}

\section{Conclusion}
In this paper the network structure of the instant messaging community of
nioki.com was investigated and demonstrated to be a scale-free network. Though
the preferential attachment was not determined, it is likely that the
Barab\'{a}si-Albert or similar model describes its evolution. Knowledge of
this structure may tell us more about social networks in the real world and
how to prevent the spread of worms on instant messaging networks.

\begin{acknowledgements}

I would like to thank the many people who helped this research be possible.
First, in the Jabber and open source community, everyone at jabber.org for
their help and suggestions including Peter Saint-Andre. Also, I
would like to thank nioki.com for working with me to get this data without
violating the privacy of their users. In particular I would like to thank
Stefan Praszalowicz. Finally, those in the physics and computer science
community  who helped with suggestions or feedback including
Albert-L\'{a}szl\'{o} Barab\'{a}si, Hawoong Jeong, Colin Steele, Steven
Strogatz, Duncan Watts, Bryan Wright, and Mark Newman for his help with my
algorithms. Finally, I would like to acknowledge the generous help of David
Smith, my advisor, under whom I completed this project. \end{acknowledgements}

\appendix
\section{Nioki.com User Privacy Considerations}
Of the utmost importance was protecting the privacy of the users of the
nioki.com instant messaging network that was researched. Here all privacy
precautions taken are outlined and explained in detail.

The data received by the reseacher of this paper was in the most raw form
possible. The data from nioki.com was prepared so that all users and their
contacts were anonymized as numbers. Users and their contacts were then
matched up by matching a user number with a contact number. The researcher did
not receive any user names, emails, IP addresses, geographical locations,
personal information or activity information, or any other data that could
allow him to either determine the identity of any given user or extrapolate
anything about a user's activity on nioki.com or the Internet at large. It
would have been impossible for the researcher to determine any direct personal
information or identity about anyone from this raw data.

The data was only in the hands of the researcher at all times and was not
given to any collaborators, published publicly or distributed in raw form, or
sold for profit. The data was statistically analyzed in aggregate and
therefore no information about any specific users could be extrapolated. A
rough metaphor of this experiment would be analyzing census data for a town.
Aggregate patterns will emerge but no specific information about individual
inhabitants can be gleamed from the data. In order to further protect privacy,
this researcher cannot distribute the raw data to any others interested, even
for research purposes. Such requests must be made directly to nioki.com.

If there are any other questions or considerations regarding the privacy of
this research, please direct them to the email at the top of the paper.

\end{document}